\documentclass[%preprint,           % Format : preprint, twocolumn
               twocolumn,
               noshowpacs,          % Pacs : showpacs, noshowpacs
               preprintnumbers,     % Preprint: preprintnumbers,
               aps,                 % Society: ...
               prd,                 % Journal Style : pra, prb, prc, prd, pre,
               a4paper,             % Size : a4paper, ...
               superscriptaddress,  % Affiliation (Title) : groupedaddress,
               nofootinbib,         % Footnote: footinbib, nofootinbib
               tightenlines,        % Remove additional spaces in a line
               floats,floatfix      % Floating pictures and tables
               ]{revtex4}

\usepackage[dvips]{graphicx}
\usepackage{latexsym}
\usepackage{amsmath,amssymb}        % amssymb includes amsfonts
\usepackage[draft=false]{hyperref}

\def\be{\begin{equation}}
\def\ee{\end{equation}}
\def\bea{\begin{eqnarray}}
\def\eea{\end{eqnarray}}
\newcommand{\ud}{\mathrm{d}}

\begin{document}

\title{A New Approach to Quintessence and a Solution of Multiple Attractors}

\author{Shuang-Yong Zhou}
\affiliation{Department of Physics, Shandong University, Jinan, 250100, P. R. China\\
{\tt Email:zhousy@mail.sdu.edu.cn}}

\begin{abstract}

We take a new approach to construct Quintessential models. With this
approach, we first easily obtain a tracker solution that is
different from those discovered before and straightforwardly find a
solution of multiple attractors, i.e., a solution with more than one
attractor for a given set of parameters. Then we propose a scenario
of Quintessence where the field jumps out of the scaling attractor
to the de-Sitter-like attractor, by introducing a field whose value
changes a certain amount in a short time, leading to the current
acceleration. We also calculate the change the field needs for a
successful jump and suggest a possible mechanism that involves
spontaneous symmetry breaking to realize the sudden change of the
field value.

\end{abstract}

\maketitle

%%%%%%%%%%%%%%%%%%%%%%%%%%%%%%%%%%%%%%%%%%%%%%%%%%%%%%%%%%%%%%%%%%%%%%%
%%%%%%%%%%%%%%%%%%%%%%%%%%%%%%%%%%%%%%%%%%%%%%%%%%%%%%%%%%%%%%%%%%%%%%%

Recent observations and experiments strongly indicate that the
universe is spatially flat and currently undergoing accelerated
expansion \cite{snae,wmap,sdss}. A negative pressure energy
component, termed dark energy, is suggested to be responsible for
the acceleration. The simplest candidate for dark energy seems to be
a positive cosmological constant, which is conventionally associated
with the quantum vacuum energy. However, it is very tiny, compared
with typical particle physics scales, which is the so-called
\emph{fine-tuning problem} \cite{fine}. It also suffers the
so-called \emph{coincidence problem} \cite{coin}. Rather than
dealing directly with the dark energy a cosmological constant,
various alternative routes have been proposed, which usually invoke
dynamical scalar fields, such as Quintessence
\cite{Wetterich88,peebles,BCN99,SW99,albrecht}, Phantom
\cite{Caldwell02} and Quintom \cite{quintom}.

Quintessence invokes an evolving canonical scalar field slowly
rolling down its potential (to some extent like the inflaton which
drives the inflation in the early universe) with equation of state
$w_{\phi}>-1$. Motivated from observational data
\cite{Corasaniti:2004sz,ASSS}, the Phantom invokes a negative
kinetic energy with effective equation of state $w_{Ph}<-1$, having
led to many interesting phenomena \cite{phantom}.

Among the various Quintessential models, \emph{tracker solutions}
have attracted a lot of attention. The tracker field has an equation
of motion with attractor-like solutions in the sense that a very
wide range of initial conditions rapidly converge to a common,
cosmic evolutionary track of $\rho_{\phi}(t)$ and $w_{\phi}(t)$. The
tracking behavior with $w_{\phi}<w_{m}$ occurs when $\Gamma>1$ and
is nearly constant ($\ud (|\Gamma-1|)/\ud \ln a\ll |\Gamma-1|$),
where $\Gamma$ is defined as $VV''/V'^2$, with V the potential and
$'$ the derivative w.r.t.~the field \cite{trcksol}. It has been
found that the general inverse power-law ($V(\phi)=\sum
c_{\alpha}/\phi^{\alpha}$) and exponential
($V(\phi)=V_0\exp(1/\phi)$) potentials are tracker solutions (we
have chosen $\kappa^2=8\pi G=1$).

Another important class of Quintessential models are \emph{scaling
solutions} \cite{CLW,van,MP,NM,gong} in which the energy density of
the scalar field mimics the background fluid energy density. Namely
scaling solutions are characterized by the relation
$\rho_{\phi}\propto\rho_{m}$, whose simplest realization is the
exponential potential $V_0e^{-\mu\phi}$. As long as the scaling
solution is the dynamical attractor, for any generic initial
conditions, the field would sooner or later enter the scaling
regime, being sub-dominant during radiation and matter dominated
eras to satisfy the tight constraints from nucleosynthesis and
structure formation, thereby opening up a new line of attack on the
fine-tuning problem \cite{dy}. However, exit from the scaling regime
is needed so as to give rise to recent acceleration.

The double exponential potential \cite{BCN99,doubleexp} of the form
\begin{equation}
\label{2exp} V(\phi) = V_0 \left( e^{-\mu\phi} + e^{-\nu\phi}
\right)\,,
\end{equation}
provides a simple realization of the exit from the scaling regime.
Such potentials can arise as a result of compactifications in
superstring models. By properly choosing $\mu$, $\nu$ and initial
conditions, one term in the potential dominated over the other
before nucleosynthesis, giving rise to the scaling solution, while
the situation has reversed recently, giving rise to a de-Sitter-like
acceleration. However, whether it is possible to obtain required
values of $\mu$ and $\nu$ remains a problem. In \cite{SW99}, the
authors considered the potential
\begin{equation}
\label{coshpot} V(\phi) = V_0 \left[ \cosh(  \mu\phi) -1 \right]^n
\,,
\end{equation}
which has two interesting asymptotical regions. One of these with
($|\mu \phi| \gg 1,~\phi<0$) gives the scaling solution, while the
other with ($|\mu \phi| \ll 1$), according to virial theorem, gives
current acceleration with average equation of state $\langle
w_{\phi} \rangle=(n-1)/(n+1)$. As current data favor an equation of
state close to $-1$ \cite{wmap}, $n$ should be close to $0$, which
is mathematically viable, but seems unnatural physically. See
\cite{as99,um} for another two popular models.

On the other hand, there is an attempt to search for a solution of
two scaling regimes by coupling Quintessence to the matter
\cite{cpquin}. Nonetheless, this scenario has faced severe
challenges since it has been showed that it cannot be realized for a
vast class of scalar field Lagrangians \cite{chllg}.

In this letter, we take a new approach to construct Quintessential
models. With this approach, instead of proposing an interesting
Quintessential potential directly, we first propose a relation
between two quantities, $\Gamma$ and $\lambda$ (defined as $-V'/V$),
and then figure out the potential. First, we show that a tracker
potential which is different from that discovered before can be
easily obtained. Then we find it straightforward to get a solution
of multiple attractors, that is, a solution with more than one
attractor for a given set of parameters. In the particular case
given in this letter, we have a scaling attractor and a
de-Sitter-like attractor. We thus propose a model in which the
universe first evolves to the scaling attractor, and then, by
introducing a field whose value changes a certain amount in a short
time, the universe jumps out to the de-Sitter-like attractor to give
the current acceleration. We also calculate the change the field
needs for a successful jump and justify the introduction of this
kind of field.

%%%%%%%%%%%%%%%%%%%%%%%%%%%%%%%%%%%%%%%%%%%%%%%%%%%%%%%%%%%%%%%%%%%%%%%%%%
%%%%%%%%%%%%%%%%%%%%%%%%%%%%%%%%%%%%%%%%%%%%%%%%%%%%%%%%%%%%%%%%%%%%%%%%%%

To start, we consider the action of Quintessence ($\epsilon=1$) (or
Phantom ($\epsilon=-1$)) minimally coupled to gravity,
\begin{equation}\label{action}
    S = \int\!\ud^4x\sqrt{-g}\: [-\frac{1}{2}\epsilon(\nabla\!
\phi)^2-V(\phi)]\,,
\end{equation}
where we use the metric signature $(-,+,+,+)$ and $(\nabla\!
\phi)^2=g^{\mu\nu}\partial_{\mu}\phi\partial_{\nu}\phi$. In the flat
Friedmann-Robertson-Walker spacetime, the equation of state for the
Quintessential field $\phi$ is given by
\begin{equation}
    w_{\phi}=\frac{p_{\phi}}{\rho_{\phi}}=\frac{\epsilon\dot{\phi}^2 -
    2V(\phi)}{\epsilon\dot{\phi}^2 + 2V(\phi)}\,.
\end{equation}
The variation of the action (\ref{action}) with respect to $\phi$
gives
\begin{equation}\label{eom}
    \epsilon\ddot{\phi}+3\epsilon H\dot{\phi}+V'=0\,.
\end{equation}

Since we carry out cosmological dynamics of the Quintessential field
$\phi$ in the presence of a barotropic fluid whose equation of state
is given by $w_m=p_m/\rho_m$ (in this paper, we assume that $w_m$ is
constant), Einstein equations reduce to
\begin{eqnarray}
\label{fdme}
    && H^2 = \frac13\; [ \frac12 \epsilon \dot{\phi}^2 + V(\phi)
       + \rho_m ]\,, \\
\label{fdme1}
    && \dot{H} = -\frac12\;[ \epsilon
       \dot{\phi}^2+(1+w_m)\rho_m ]\,.
\end{eqnarray}
Introducing the following dimensionless variables
\begin{eqnarray}
    && x \equiv \frac{\dot{\phi}}{\sqrt{6}H}\,,~~
        y \equiv \frac{\sqrt{V}}{\sqrt{3}H}\,,\nonumber \\
\label{lamGam}
    && \lambda \equiv -\frac{V'}{V}\,,~~\;
       \Gamma \equiv \frac{VV''}{V'^2}\,,
\end{eqnarray}
Eq.~(\ref{eom}), (\ref{fdme}), (\ref{fdme1}) can be recast in the
following form \cite{CLW,Ng,dy}:
\begin{eqnarray}
\label{autoquin1} \hspace*{-1.5em} \frac{\ud x}{\ud N} &=&
-3x+\frac{\sqrt{6}}{2} \epsilon \lambda y^2 \nonumber \\
& & +\frac32 x[(1-w_m)\epsilon x^2 +(1+w_m)(1-y^2)]\,, \\
\label{autoquin2}
\hspace*{-1.5em} \frac{\ud y}{\ud N} &=&
-\frac{\sqrt{6}}{2}\lambda xy \nonumber \\
& & +\frac32 y[(1-w_m)\epsilon x^2 +(1+w_m)(1-y^2)]\,, \\
\label{autoquin3}
\hspace*{-1.5em}\frac{\ud \lambda}{\ud N} &=&
-\sqrt{6} \lambda^2 (\Gamma-1)x\,,
\end{eqnarray}
where $N=\ln a$ ($a$ is the scale factor), together with a
constraint equation
\begin{equation}
\label{confine} \epsilon x^2+y^2+\frac{\rho_{m}}{3H^2}=1\,.
\end{equation}
The equation of state $w_{\phi}$ and the fraction of the energy
density $\Omega_{\phi}$ for the field $\phi$ are, respectively,
\bea
\label{wphiquin}
& & w_{\phi} \equiv \frac{p_{\phi}}
{\rho_{\phi}}=\frac{\epsilon x^2-y^2}
{\epsilon x^2+y^2}\,, \\
& &
\label{Omephiquin}
 \Omega_{\phi} \equiv \frac{\rho_{\phi}}{3H^2} =\epsilon x^2+y^2\,.
\eea

To warm up, we note that for many Quintessential (or Phantom)
potentials $\Gamma$ can be written as a function of $\lambda$. Let
us take the case of Phantom potential of the form
\begin{equation}
    V(\phi)=\frac{V_0}{[\cosh(\sigma\phi)]^n}
\end{equation}
for example. It is found
\be \label{phangam} \Gamma=1 + \frac1n -
\frac{n\sigma^2}{\lambda^2}\,. \ee
Substituting Eq.~(\ref{phangam}) into Eq.~(\ref{autoquin3}), we can
perform three-dimension dynamical analysis of the autonomous system.
For a barotropic fluid background, there is a unique stable fixed
point $(x=0,\;y=1,\;\lambda=0)$, which is a de-Sitter-like dominant
attractor. For the case $n=1$, it confirms the results of
\cite{SSN}. Note that we neglect the cases with $y<0$, as the system
is symmetric under the reflection $(x,y)\to(x,-y)$ and time reversal
$t\to-t$.

The direct way to get a Quintessential (or Phantom) model is to
conceive (usually fairly carefully) a potential that meets
constraints from observations and experiments. However, encouraged
by what has been showed above, let us take another route.

We note that the dynamical system (\ref{autoquin1}, \ref{autoquin2},
\ref{autoquin3}), is obviously autonomous except for $\Gamma$. In
fact, since the potential $V(\phi)$ is only a function of the field
$\phi$, by the definition (\ref{lamGam}), $\lambda$ and $\Gamma$ can
be written as
\be \lambda=P(\phi)\,,~~~~\Gamma=Q(\phi)\,. \ee
If the inverse function of $P(\phi)$ exists, then we have
\be \label{para} \Gamma=Q(P^{-1}(\lambda))\equiv f(\lambda)\,. \ee

It is noteworthy that in principle we can figure out the potential
as a function of the field $\phi$. Using the definition of $\lambda$
and $\Gamma$, Eq.~(\ref{para}) can be can rewritten as
\be V''=\frac{{V'}^2}{V}f(-\frac{V'}{V})\equiv F(V,V')\,. \ee
Let $h=V'$, then we get
\be \frac{\ud h}{\ud V}=\frac1h F(V,h)\,. \ee
Having figured out $h(V)$, we can solve $V'(\phi)=h(V(\phi))$ to
obtain the potential $V(\phi)$. Thus we can perform three-dimension
dynamical analysis of the system (\ref{autoquin1}, \ref{autoquin2},
\ref{autoquin3}) with a fairly large amount of potentials beyond the
exponential case where the dynamical system reduces to two-dimension
autonomous system.

In the viewpoint of $\Gamma$ as a function of $\lambda$ and
considering the powerful theorem presented by \cite{trcksol}, it is
easy to obtain a tracker field. As an example, we write $\Gamma$ as
\be \Gamma=1+\frac{\alpha}{\lambda^2}\,, \ee
which can be solved to give the potential
\be V(\phi)=V_0e^{\frac{\alpha}{2}\phi(\phi+\beta)}\,, \ee
where $V_0(>0)$ and $\beta$ are integral constants. We note that it
is different from the general inverse power-law ($V(\phi)=\sum
c_{\alpha}/\phi^{\alpha}$) or exponential
($V(\phi)=V_0\exp(1/\phi)$) potentials.

Obviously, $\Gamma>1$ if $\alpha>0$. To confirm this is a real
tracker solution, we perform the condition $\ud (|\Gamma-1|)/\ud
N\ll |\Gamma-1|$, and we get
\be \label{trck1} |2\frac{\ud \lambda/\ud N}{\lambda}|\ll 1\,. \ee
Substituting Eq.~(\ref{autoquin3}) for Eq.~(\ref{trck1}), we obtain
\be |\alpha x|\ll|\lambda|\,. \ee
Considering the tracking condition $|\lambda|\sim|1/x|$
\cite{trcksol}, we finally get
\be \frac{\alpha}{\lambda^2}\ll 1\,. \ee

\begin{figure}
\includegraphics[height=2.3in,width=3.3in]{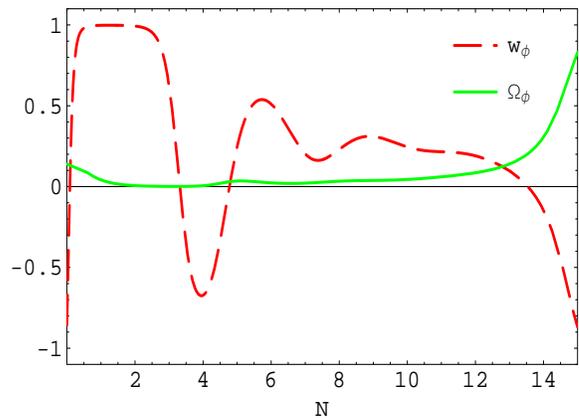}
\caption{Evolution of $w_{\phi}$ (red dashed line) and
$\Omega_{\phi}$ (green solid line) of $\Gamma=1+\alpha/\lambda^2$
($V(\phi)=V_0e^{\alpha\phi(\phi+\beta)/2}$, $\alpha$ is chosen as
$2.8$) with respect to $N=\ln a$ in the background fluid with
$w_m=1/3$. We choose initial conditions as $x_i=0.1$, $y_i=0.36$ and
$\lambda_i=17.8$. For simplicity we have neglected the
matter-dominated era.} \label{tracker}
\end{figure}

Tracking behaviors exist for a wide range parameters and initial
conditions, which solves the fine-tuning problem. However, due to
the $w$-$\Omega$ relation that alleviates the coincidence problem,
it is difficult to obtain current equation of state $w_0<-0.8$. A
numerical solution of the cosmic dynamical evolution with tracking
behavior, where for simplicity we have neglected the
matter-dominated era, is given by Fig.~\ref{tracker}.

%%%%%%%%%%%%%%%%%%%%%%%%%
%%%%%%%%%%%%%%%%%%%%%%%%%

%%%%%%%%%%%%%%%%%%%%%%%%%%%%%%%%%%%%%%%%%%%%%%%%%%%%%%%%%%%%%%%%%%%%%%%%%%
%%%%%%%%%%%%%%%%%%%%%%%%%%%%%%%%%%%%%%%%%%%%%%%%%%%%%%%%%%%%%%%%%%%%%%%%%%

Having witnessed the utility of this approach, we would like to go
further. As showed below, we find it straightforward to parameter
$\Gamma$ as function of $\lambda$ to get a solution of multiple
attractors, i.e., a solution with more than one attractor for a
given set of parameters. In the particular case given below, we have
a scaling attractor and a de-Sitter-like attractor and it is worth
noting that two scaling solutions are problematic \cite{chllg}. Thus
we are encouraged to consider a scenario that the initial conditions
of the cosmic scalar field are in the basin of a scaling solution
and first the field evolves toward the scaling attractor. Then
recently, the field \emph{jumps} out to the basin of a
de-Sitter-like dominant attractor, giving rise to the current
acceleration. In this scenario, the mechanism of exit from the
scaling regime is different from those mentioned in the
introduction, which typically invoke fairly carefully conceived
potentials with two asymptotical behaviors corresponding to the
scaling case and the de-Sitter-like case respectively. Therefore
attractors in those models are not exact. On the contrary, the two
attractors considered below are exact and we suggest some other
physical reason to realize the exit from the scaling regime, rather
than connect the two interesting regimes with more or less contrived
potentials. The physical reason is formulated as the sudden change
of the field value.

Considering Eq.~(\ref{autoquin3}), we parameter $\Gamma$ as
\be \Gamma=1+1/\beta+\frac{\alpha}{\lambda}\,. \ee
There are at least the following two fixed points:
\begin{itemize}
    \item Point(\emph{a}):\\
    ($x=-\sqrt{6}\alpha\beta,\;y=\sqrt{1-\alpha^2\beta^2/6},\;
          \lambda=-\alpha\beta$) is a
          de-Sitter-like dominant attractor, in which
          $w_{\phi}=-1+\alpha^2\beta^2/3$.
          The eigenvalues of the Jacobi matrix of the dynamical system
          are
          \begin{eqnarray}
          &&\mu_1=-\alpha^2\beta\,, \nonumber \\
          &&\mu_2=-3+\frac{\alpha^2\beta^2}{2}\,,  \nonumber \\
          &&\mu_3=-3(1+w_m)+\alpha^2\beta^2\,.  \nonumber
          \end{eqnarray}
          It exists if $\alpha^2\beta^2<6$ and is stable if $\alpha^2\beta^2<3(1+w_m)$ and
          $\beta>0$.
    \item Point(\emph{b}):\\
    ($x=\!-\sqrt3(1\!+\!w_m)/\sqrt2\alpha\beta,\;y=\!\!\sqrt{3(1\!-\!w_m^2)/2\alpha^2\beta^2},\linebreak[4]
          \lambda=-\alpha\beta$) is a scaling
          attractor, in which $w_{\phi}=w_m$ and $\Omega_{\phi}=3(1+w_m)/\alpha^2\beta^2$.
          The eigenvalues of the Jacobi matrix are
          \begin{eqnarray}
          && \mu_1=-\frac{3(1+w_m)}{\beta}\,, \nonumber \\
          && \mu_{2,3}=-\frac34(1-w_m)(1\pm\sqrt{\frac{-7-9w_m+24(1+w_m)^2}{(1-w_m)\alpha^2\beta^2}})\,.  \nonumber
          \end{eqnarray}
          It exists if $\alpha^2\beta^2>3(1+w_m)$ and is stable if $\alpha^2\beta^2>3(1+w_m)$ and
          $\beta>0$.
\end{itemize}
%

%%%%%%%%%%%%%%%%%%%%%%%%%
\begin{figure}
\includegraphics[height=2.3in,width=3.3in]{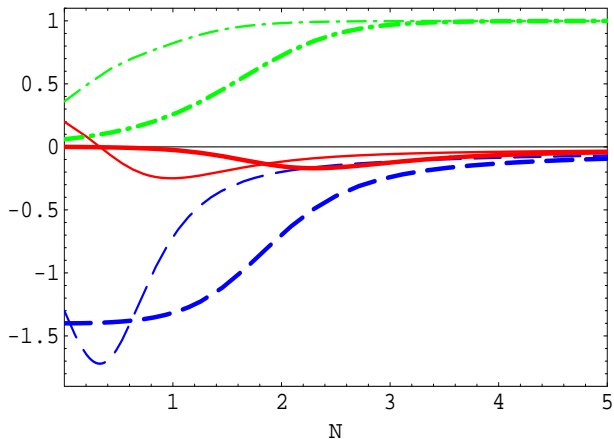}
\caption{Evolution of $x$ (red solid line), $y$ (green dot-dashed
line), $\lambda$ (blue dashed line) of
$\Gamma=1+1/\beta+\alpha/\lambda$ with respect to $N=\ln a$ in the
background fluid with $w_m=0$. $\alpha$ is chosen as $-2.6$, $\beta$
chosen as $2$. We choose initial conditions as $x_i=0.2$, $y_i=0.36$
and $\lambda_i=-1.3$ for the thin lines, and $x_i=0$, $y_i=0.06$ and
$\lambda_i=-1.4$ for the thick lines. Note that the attractor is a
stable spiral.} \label{mulattr}
\end{figure}
%%%%%%%%%%%%%%%%%%%%%%%%%

For the stabilities of the fixed points \emph{a} or \emph{b}, we
choose $\beta>0$. We note that the fixed points \emph{a} and
\emph{b} are typical for general scaling solutions and cannot be
both stable for a given set of parameters \cite{st,dy}.

Besides, we find that the de-Sitter-like dominant fixed point
($x=0,\;y=1,\;\lambda=0$) is stable, i.e., a de-Sitter-like dominant
attractor which we denote as Point(\emph{c}). It cannot be simply
seen from the eigenvalues of the Jacobi matrix
($\mu_1=-3(1+w_m),\;\mu_2=-3,\;\mu_3=0$), since $\mu_3=0$. However,
it can be seen from numerical simulation of the dynamical system,
see Fig.~\ref{mulattr} for a working example. It is found that when
$\alpha<0$ $(\alpha>0)$, the region of $\lambda<0$ $(\lambda>0)$ in
the phase space of the dynamical system is the basin of the
attractor ($x=0,\;y=1,\;\lambda=0$), while the rest region is the
basin of the attractor with $\lambda=-\alpha\beta$, i.e., (\emph{a})
or (\emph{b}). This is desirable, since the basins are divided by a
plane of equal $\lambda$ in the phase space, having nothing to do
with $x$ and $y$. Thus the initial values of $x$ and $y$ can be
arbitrary.

The corresponding potential of this case is
\be V(\phi)=\frac{V_0}{(\eta+e^{-\alpha\phi})^{\beta}}\,, \ee
where $V_0(>0)$ and $\eta$ are integral constants. For the stability
of the potential, that is, the potential should be bounded below, we
should choose $\beta$ as $2,4,6,...$ and to obtain interesting cases
we choose $\eta<0$.

Now we come to consider the scenario that the field exits the
scaling regime to the de-Sitter-like regime due to a sudden change
of the field value. To this end, we conceive $\phi$ as
\be \label{ansatz} \phi=f(t)\varphi\,, \ee
with
\be \label{chg} f(t)=\left\{
\begin{array}{ll}
1\quad\;\; t<t_j & \\
f_j\quad\, t\geq t_j\,,
\end{array}\right.
\ee
where $t$ is the cosmic time and $f_j$ is a constant with the
subscript $j$ representing some recent time when the field jumps.
Note that $f(t)$ is not necessarily of the form above, but it should
have a certain amount of change of its value in a short time so that
the field will not evolve back to the scaling attractor. We will
first calculate the change the field needs for a successful jump and
then suggest a possible mechanism to realize the sudden change of
the field value.

In order to calculate the change the field needs to realize the
jump, we choose, without losing generality, $\alpha>0$ so that the
region of $\lambda>0$ is the basin of the de-Sitter-like dominant
attractor (\emph{c}). To meet the constraints from nucleosynthesis
and structure formation, we require that the Quintessential field
have well scaled with the background before nucleosynthesis. So we
choose $\alpha^2\beta^2>20$ for (${\Omega_{\phi}} ({T \sim 1 {\rm
MeV}}) < 0.2$). At some recent time just before the jump the field
$\phi=\phi_j=\varphi_j$ and then $\phi$ rapidly changes from
$\phi_j$ to $\phi_j+\delta \phi$ (or from $\varphi_j$ to
$f_j\varphi_j$). For jumping from the basin of the scaling attractor
(\emph{b}) to that of the de-Sitter-like dominant attractor
(\emph{c}) happening, we have
\be \delta\phi>-\frac1{\alpha}\ln
(-\eta)-\phi_j=\frac1{\alpha}\ln{\frac{\lambda_j}{\lambda_j+\alpha\beta}}
\ee
where $\lambda_j=-\alpha\beta/(1+\eta e^{\alpha\phi_j})$, or
\be \label{sgm} f_j\varphi_j>-\frac1{\alpha}\ln (-\eta)\,. \ee

Now we shall justify the introduction of the field whose value has a
sudden change. Below, we suggest a possible mechanism, which resorts
to spontaneous symmetry breaking, to realize the sudden change of
the field value. We first note that scalar fields are ubiquitous in
supersymmetric theory of particle physics. Thus it is reasonable to
assume that a few of them are relevant to the cosmic evolution.
Considering $\phi$ as an effective field, we involve three fields
with the Lagrangian
\bea \label{hyb} \mathcal{L}&=&-\frac12f^2(\sigma)
g^{\mu\nu}\partial_{\mu}\varphi\partial_{\nu}\varphi
-\frac{V_0}{(\eta+e^{-\alpha\varphi f(\sigma)})^{\beta}} \nonumber\\
&&-\frac12
g^{\mu\nu}\partial_{\mu}\sigma\partial_{\nu}\sigma-\frac12
g^{\mu\nu}\partial_{\mu}\theta\partial_{\nu}\theta-V(\sigma,\theta)\,,
\eea
where
\bea
V(\sigma,\theta)&=&V'_0-\frac12m_{\sigma}^2\sigma^2+\frac14\lambda\sigma^4+\frac12m^2\theta^2
+\frac12\lambda'\sigma^2\theta^2 \nonumber\\
&=&\frac14\lambda(M^2-\sigma^2)^2+\frac12m^2\theta^2+\frac12\lambda'\sigma^2\theta^2\,.
\eea
We conceive $f(\sigma)$ as
\be \label{chg} f(\sigma)=\left\{
\begin{array}{ll}
1\quad\;\; \sigma^2<\sigma_s^2 & \\
f_j\quad\, \sigma^2\geq \sigma_s^2\,,
\end{array}\right.
\ee
so that $\varphi$ decouples from $\sigma$ and $\theta$ except at the
points $\sigma=\pm\sigma_s$ (they surely couple to each other
through Friedmann equation; nevertheless, when the radiation or
matter dominates the universe, the coupling through gravity is
neglectable.). Since $f(\sigma)$ is more reasonable to be a
continuous function, a better choice of $f(\sigma)$ might be
\be f(\sigma)=\frac{1+f_j}2-\frac{1-f_j}2\tanh
[a(\sigma^2-\sigma_s^2)]\,,~~a\gg1\,.\ee

We note that $V(\sigma,\theta)$ is famous for its realization of
Hybrid inflation models \cite{hyb}. In these models, first, $\sigma$
is held at the origin, with $\theta$ slow rolling down the
potential, giving the inflation, and then, when $\theta$ rolls down
a critical value $\theta_c$, $\sigma$ is destabilized and quickly
rolls down from $0$ to $\pm M$, ending the inflation. Comparing two
ways of writing the potential, we obtain
\bea m_{\sigma}^2&=&\lambda M^2\,,\\
V'_0&=&\frac14\lambda M^4\,. \eea
And the critical value of $\theta$ is
\be \theta^2_c=m_{\sigma}^2/\lambda'=\lambda M^2/\lambda'\,. \ee
For this potential to be viable for current purpose, $\theta$ does
not necessarily slow roll. Yet we do need $\sigma$ quickly roll down
from $0$ to $¡ÀM$ after $\theta$ rolls down $\theta_c$, which
implies
\be m^2_{\sigma}\gg V'_0\,, \ee
and we require $0<\sigma_s^2<M^2$ so that when $\sigma$ rolls down
from $0$ to $\pm M$, $f(\sigma)$ changes from $1$ to $f_j$.

Note that when the field $\phi$ jumps to the de-Sitter-like regime,
$V(\phi=\varphi f(\sigma))$ will increase and the kinetic term of
$\varphi$ will also change. At the same time, $V(\sigma,\theta)$
should decrease so as to vanish when Quintessence begins to dominate
the universe. For this scenario to be viable, we require that the
decrease of $V(\sigma,\theta)$ be larger than the increase of
$V(\phi=\varphi f(\sigma))$ and the kinetic term (note that when
$|f_j|<1$, the kinetic term will decrease; however, it is easy to
show that the decline of the kinetic term in this case is small,
compared to the increase of the $V(\phi=\varphi f(\sigma))$.). One
may worry that this might spoil the analysis of the dynamics of
$\phi$ above, as this require the energy associated with $\sigma$
and $\theta$ to be comparable with that associated with $\varphi$
around the jump point. However, we argue that it will not, because
$\varphi$ almost decouples to $\sigma$ and $\theta$, and the energy
associated with $\sigma$ and $\theta$ is only comparable with that
associated with $\varphi$ when the radiation or matter dominates the
universe and vanishes when the dark energy begins to dominate the
universe.

%%%%%%%%%%%%%%%%%%%%%%%%%%%%%%%%%%%%%%%%%%%%%%%%%%%%%%%%%%%%%%%%%%%%%%%%%%%%%
%%%%%%%%%%%%%%%%%%%%%%%%%%%%%%%%%%%%%%%%%%%%%%%%%%%%%%%%%%%%%%%%%%%%%%%%%%%%%

In summary, we suggest a new approach to construct Quintessential
dark energy models, with which we first propose a relation
$\Gamma=f(\lambda)$ between $\Gamma=VV''/(V')^2$ and
$\lambda=-V'/V$, and then figure out the potential $V(\phi)$. It is
showed that a tracker solution that is different from those
discovered before can be easily obtained and a solution of multiple
attractors is also found straightforwardly. Then we suggest a
scenario that the initial conditions of the cosmic scalar field are
in the basin of a scaling attractor and first, the field evolves
toward the scaling attractor and then recently, the field jumps out
to the basin of a de-Sitter-like dominant attractor, giving rise to
current acceleration. For this scenario to be realized, we introduce
a field whose value changes a certain amount in a short time. Then
we calculate the change the field needs for a successful jump and
invoke a mechanism that is similar to the case of Hybrid inflation
to justify the introduction of this kind of field.

%%%%%%%%%%%%%%%%%%%%%%%%%%%%%%%%%%%%%%%%%%%%%%%%%%%%%%%%%%%%%%%%%%%%%%%%%%%%%%
%%%%%%%%%%%%%%%%%%%%%%%%%%%%%%%%%%%%%%%%%%%%%%%%%%%%%%%%%%%%%%%%%%%%%%%%%%%%%%

We thank Yun-Song Piao, Yi Wang, Zuo-Tang Liang, Jian-Hua Gao and Ye
Chen for useful discussions.

%%%%%%%%%%%%%%%%%%%%%%%%%%%%%%%%%%%%%%%%%%%%%%%%%%%%%%%%%%%%%%%%%%%%%%%%%%%%%%
%%%%%%%%%%%%%%%%%%%%%%%%%%%%%%%%%%%%%%%%%%%%%%%%%%%%%%%%%%%%%%%%%%%%%%%%%%%%%%

\end{document}